\documentclass[12pt]{article}

\usepackage{amssymb,amsmath}                                           % Standard math packages
\usepackage{bbm}                                                       % Fancy blackborld bold
\usepackage{tensor}                                                    % For fancy indices
\usepackage[usenames,dvipsnames]{color}                                % For colors (for hyperlinks)
\usepackage{setspace}                                                  % For spacing,  must come BEFORE `hyperref' package.
\usepackage{hyperref}                                                  % For hyperlinks
\usepackage[left= 1in,right=1in,top=1in,bottom=1in]{geometry}          % For margin settings
\usepackage{comment}                                                   % For comment environment
\usepackage{authblk}                                                   % For author affiliations, etc...
\usepackage{graphicx}                                                  % For ``includegraphics''
\usepackage{caption}                                                    % For fancy captions and subcaptions.
\usepackage{subcaption}

\hypersetup{                 
    colorlinks=true,         % false: boxed links; true: colored links, false is default
    linkcolor=blue,          % color of internal links, red is default
    citecolor=red,           % color of links to bibliography, 'green' is default
    urlcolor=Violet          % color of external links, cyan is default
}

\let\oldmarginpar\marginpar
\renewcommand\marginpar[1]{\oldmarginpar{\color{red}\raggedright\scriptsize #1}}
\newcommand{\pb}[2]{\ensuremath{\lf\{#1,#2 \rt\}}}

\newcommand{\diby}[2]{\ensuremath{\frac{\partial #1}{\partial #2}}}

\newcommand{\w}[2]{\ensuremath{\omega\indices{^{#1}_{#2}}}}
\newcommand{\sfrac}[2]{\ensuremath{{\textstyle \frac{#1}{#2}}}}

\def\lf {\ensuremath{\left}}
\def\rt {\ensuremath{\right}}
\def\de {{\rm d}}
\def\De {{\rm D}}
\def\in {\ensuremath{_{\text{in}}}}
\def\out {\ensuremath{_{\text{out}}}}
\def\outin {\ensuremath{_{\text{out/in}}}}
\def\mm {\ensuremath{\mathcal M}}
\def\ham {\ensuremath{\mathcal H}}

\title{Observing Shape in Spacetime}

\author[1,2]{{\bf Sean Gryb}\thanks{email: \href{mailto:s.gryb@hef.ru.nl}{s.gryb@hef.ru.nl}}}
\affil[1]{\footnotesize {\it Institute for Theoretical Physics, Utrecht University, Leuvenlaan 4, 3584 CE Utrecht, The~Netherlands}}
\affil[2]{{\it Radboud University Nijmegen, Institute for Mathematics, Astrophysics and Particle Physics, The~Netherlands}}

\date{\today}

\begin{document}

\maketitle

\begin{abstract}
	The notion of \emph{reference frame} is a central theoretical construct for interpreting the physical implications of spacetime diffeomorphism invariance in General Relativity. However, the alternative formulation of classical General Relativity known as \emph{Shape Dynamics} suggest that a subset of spacetime diffeomorphisms --- namely hypersurface deformations --- are, in a certain sense, dual to spatial conformal (or Weyl) invariance. Moreover, holographic gauge/gravity dualities suggest that bulk spacetime diffeomorphism invariance can be \emph{replaced} by the properties of boundary CFTs. How can these new frameworks be compatible with the traditional notion of reference frame so fundamental to our interpretation of General Relativity? In this paper, we address this question by investigating the classical case of maximally symmetric spacetimes with a positive cosmological constant. We find that it is possible to define a notion of \emph{Shape Observer} that represents a conformal reference frame that is dual to the notion of inertial reference frame in spacetime. We then provide a precise dictionary relating the two notions. These Shape Observers are holographic in the sense that they are defined on the asymptotic conformal boundaries of spacetime but know about bulk physics. This leads to a first principles derivation of an exact classical holographic correspondence that can easily be generalized to more complicated situations and may lead to insights regarding the interpretation of the conformal invariance manifest in Shape Dynamics.
\end{abstract}
\clearpage

\tableofcontents
\clearpage
 
\section{Introduction}

\subsection{Coordinate Invariance and the Equivalence of Frames}

Before getting to the primary purpose of this paper, let me first try to describe the main problem to which the key results of this paper are meant to be addressed. Readers less interested in philosophical motivations can skip this section and proceed directly to Section~\ref{sec:intro SO} where the main results are introduced. In brief, the problem we will be concerned with regards the precise content and, ultimately, interchangeability of the assumptions that are made when making inferences about `when' and `where' some event, not in one's immediate neighbourhood, take place. In General Relativity, these assumptions are intimately tied with the notion of a spacetime and spacetime diffeomorphism invariance, which is a gauge symmetry of the theory. Consequently, coordinate values are taken to be `pure gauge' within the formalism. Indeed, it is well-known that Einstein placed primary importance on this fact \cite{einstein:gen_rel}, which he considered a key foundational principle of the theory. But how are we to understand the physical significance of spacetime coordinate invariance in General Relativity?

On the one hand, there is certainly something \emph{real} about the hand of a clock pointing to 3 (or 5 or whatever) just as there is something \emph{real} in the observation that an object has a width equal to 15 tick marks of a ruler. On the other hand, one always has the freedom to label one's clock differently --- say by swapping the 3 with a 5 --- or change the increments of the tick marks of one's ruler. As Kretschmann pointed out long ago \cite{kretschmann:gen_cov}, these choices shouldn't have any impact on objective facts about events. This objection raises serious doubts as to Einstein's intuition regarding the principle of general covariance and its connection with coordinate invariance. There is, however, a more precise way to bring Einstein's intuition into a slightly more tenable position, which we will now briefly describe.

The key insight is to recognize the difference between the coordinate values, $x^\mu$, which serve as arbitrary labels and have no physical significance, and the readings of real physical clocks and rods, which are represented by the values of independent scalar fields, say $X^a(x^\mu)$ (where $a = 1, \hdots, 4)$ , within the theory and do have physical significance.\footnote{In fact, it can easily be seen (see \cite{Westman:2007yx}) that identifying the $x^\mu$'s with physical observables leads to an under-determination problem in the equations of motion due to the coordinate invariance of the equations of motion.} It is what the coordinate invariance of the theory implies for the properties of these scalar fields --- i.e., that no particular configuration of $X^a$'s should be preferred over any other by the formalism --- that constitutes the physical content of this symmetry. 

To formulate this position more concretely, we can define the useful theoretical construct of a \emph{reference frame}. Physically, a reference frame represents a system of ideal (non-colliding) clocks and rods that is large enough and dense enough to measure times and locations of events throughout some spatial region of interest. These ideal clocks and rods are assumed to be identically constructed so that they can measure time and position accurately and reproducibly throughout space. Moreover, a canonical clock synchronisation procedure must be specified --- perhaps by sending light signals back and forth between clocks and making assumptions about the propagation of these light signals --- allowing a local observer to \emph{infer} the readings and positions of distant clocks. Formally, the worldlines of these ideal clocks and rods are represented by a time-like congruence in a spacetime manifold $\mathcal M$. This congruence defines a time-like vector field $u^\mu$ through the tangents to its worldlines. In Appendix~\ref{apx:ref frames}, we show explicitly how it is possible to extract physical clock and rod readings, $X^a(x^\mu)$, from a suitable time-like vector field $u^\mu(x^\mu)$ and the conditions that need to be satisfied by $u^\mu$. An excellent example implementing this idea is given by the Gaussian reference fluids considered by Kucha\v r and Torre in \cite{Kuchar:1990vy}.

We can now distinguish two distinct notions of ``coordinate transformation''. The first --- understood as a \emph{passive} transformation --- entails keeping the \emph{same} physical clocks and rods while transforming the coordinates labels only, which corresponds to using \emph{different} conventions for making readings of the clocks and rods. The second --- understood as an active transformation --- entails using identically constructed but physically \emph{distinct} clocks and rods in different states of motion while keeping the \emph{same} conventions, i.e., the same coordinate labels, for making readings. Clearly, the first notion corresponds to Kretschmann's notion and has no physical content while the second is more significant and does have physical consequences. That General Relativity exhibits an invariance under the second notion of active coordinates transformations is guaranteed by two necessary ingredients: the first is the variational principle of the theory, which varies over all possible configurations of both matter fields and spacetime metrics, and the second is the coordinate invariance of the action. It is the combination of these two properties of General Relativity that guarantee that no reference frame is favoured \emph{a priori} by the formalism.

To understand this, consider the case where the reference frame is represented by a geodesic congruence given, for example, by a pressureless dust. These reference frames are preferred by the formalism because of the equivalence principle: locally they are indistinguishable from the inertial frames of Special Relativity. However, as pointed out by \cite{Dieks:frame_equiv}, these frames are only specified once the spacetime geometry has been determined by solving the variational principle of the Einstein--Hilbert action, which varies over all possible matter and geometry configurations. Furthermore, because the variational principle itself is diffeomorphism invariant, no choice of arbitrary labels can be used to single out a preferred reference frame. Thus, the reference frames are not signed out by nomological law-like restrictions (the variational principle) but rather contingent circumstance (the initial and boundary conditions of the variational principle). It is in this way that General Relativity can be seen to not privilege any particular choice of reference frame.

The purpose of this elaborate discussion is twofold. First, it is to point out the importance of the notion of reference frame when formulating two key foundational principles of General Relativity, namely: the equivalence principle and the physical principle behind general coordinate invariance. Secondly, it is to point out the number of assumptions and conventions that must be used to make these principles precise. Crucially, it is these assumptions and conventions that frame our understanding of how a local observer infers clock and rod readings for distant events. Given the number of assumptions and conventions of this construction, it would not be a surprise if there existed a completely \emph{different} set of assumptions and conventions that could be used to make \emph{different}, but physically equivalent, coordinate assignments to distant events. This would lead to a very different kind of reference frame than that entailed by General Relativity. Indeed, there would even be no reason to expect that the symmetry group defining the equivalence class of \emph{new} reference frames should be that of active spacetime diffeomorphisms. Nevertheless, it could still be possible that this new notion of reference frame could define the same physics as the old one.

For this to be possible, what is needed is a way to map between the assumptions and conventions of the old notion of reference frame to the assumptions and conventions of the new notion of reference frame. Furthermore, this map should induce an isomorphism between the coordinate assignments of the old formalism to those of the new one. This ensures that events described by the old references frames are completely and uniquely described by the new ones. An immediate consequence of this is that the symmetry group defining the equivalence class of old frames should be isomorphic to the new symmetry group defining an equivalence class of new frames. This is because the equivalence class itself has a physical interpretation within the formalism: it represents the space of \emph{potential} physical (albeit hypothetical) reference frames. If a map between the old and new symmetry groups could be understood explicitly, then the difference between the two formalisms would be purely interpretive because the formalisms would only differ with regards to the conventions a local observer uses to make inferences about the coordinates values of events of distant observers.

That such a new notion of reference frame exists is hinted at by a new theory of gravitation known as \emph{Shape Dynamics} \cite{gryb:shape_dyn,Gomes:linking_paper,JuliansReview,gryb:thesis}. In this theory, the local symmetry group is not that of spacetime diffeomorphisms but, rather, that of \emph{foliation-preserving} (spatial) diffeomorphisms and spatial local conformal (or Weyl) transformations. Shape Dynamics, thus, has a preferred notion of simultaneity but a completely arbitrary notion of local scale. Despite this profound difference from General Relativity, a vast number of solutions of Shape Dynamics can be shown to be solutions of the Einstein equations when expressed using this preferred notion of simultaneity. It would appear that the principle of spacetime general covariance can be interchanged for the principle of spatial general covariance and local scale invariance. If one accepts this, then it must also follow that the notion of reference frame --- so intimately tied to the physical principle behind spacetime general covariance --- must be interchangeable with a new notion of reference frame to be tied to the new physical principles behind by spatial general covariance and local scale invariance. In fact, without such a new notion of reference frame, the statement that the two formalisms are ``dual'' is put strongly into question given the key role played by reference frames (and their equivalence) in the foundational principles of General Relativity.

Unfortunately, the ``duality'' between Shape Dynamics and General Relativity is currently expressed only in terms of a gauge-theory inspired language of equivalence of Dirac observables. Concretely (see \cite{gryb:shape_dyn,Gomes:linking_paper} for details), specific gauge fixings of both theories lead to the same trajectory on phase space given some valid initial data. However, the foundational principles and, more specifically, the physical significance of the local symmetries of the action are \emph{very} different in General Relativity compared with standard gauge theories like Yang--Mills theory. In Yang--Mills theory, for example, the notion of reference frame is completely unnecessary for understanding the significance of the gauge invariance of the theory, which is purely associated with an indeterminacy in the mathematical formalism used to model a physical system. It would appear that, without an explicit construction of a Shape Dynamics' reference frame and a map between this and a reference frame in General Relativity, we have a strong reason to question whether there is a strict duality between both formalisms.

\subsection{Shape Observers}\label{sec:intro SO}

We are now in a position to properly motivate the intentions of the current paper. Our goal here will be to explore a bottom-up approach to constructing a new notion of reference frame, which we will call a \emph{Shape Observer}, that is intended to provide an appropriate notion of observer for Shape Dynamics. We will see that requiring a strict notion of equivalence between these Shape Observers and the reference frames of General Relativity leads to a \emph{holographic} correspondence in the sense of \cite{Hooft:1993gx,Susskind:1994vu}.

To establish a notion of Shape Observer, we will consider the comments of the previous section and make use of the well-known isomorphism between the global isometry group of de~Sitter (dS) spacetime in $D+1$ dimensions, namely $SO(D+1,1)$, and the conformal group $Conf(D)$ in $D$ dimensional space.\footnote{We could have, in contrast, considered the isometry between the anti-de~Sitter group and $Conf(D-1,1)$. The results would be quantitatively similar but the physical interpretation would not be in terms of shapes.} This will require that we consider spacetimes that are such that local inertial observers see the dS metric rather than the Minkowski metric, which is perhaps the case more relevant to experimental observations. For simplicity, we will, in addition, restrict our attention to the maximally symmetric case. Consequently, we will consider only the equivalence class of (global) inertial observers in dS spacetime. Since these are related mathematically by the isometry group of dS, we can use the isormorphism mentioned above to construct a notion of reference frame where equivalence classes of the new frames are related by the transformations of the spatial conformal group. Because these new frames are equivalent with regards to the scale of some collection of particles, the only physically relevant information is contained in the scale-invariant `shape' of the system. This will define for us the notion of Shape Observer. Many of the mathematical structures enable us to define these Shape Observers are discussed in \cite{Wise:2013uma}.

The main difficulty posed in using the isomorphism mentioned above to construct a notion of Shape Observer from the inertial observers of dS is that, while the isomorphism exists at the abstract group level, finding an explicit map between the \emph{actions} of the dS group on spacetime and the conformal group on the Euclidean plane is much more difficult. In particular, the role of time (which drops out of the action of the conformal group on the plane) must be considered very carefully. The problem originates from the fact that the notion of \emph{instantaneous} configuration, which is necessary to define a shape, is ill-defined in the spacetime picture because of relativity of simultaneity: different inertial observers won't agree on which instant should be used to define a shape.

To deal with this, we will make use of the properties of the conformal boundaries of dS to establish a precise notion of what we will call \emph{shape freezing}. We then make use of this shape freezing to use the bulk action for free particles in dS to define the Hamilton--Jacobi function for a conformally invariant and reparametrization invariant particle model on the Euclidean plane. The reparametrization invariance is a consequence of the inferred role of duration in the dual description. This gives us a holographically defined dual description of a system of free particles in dS spacetime in terms of a conformally invariant system of particles on the plane.

The holographic nature of the definition of Shape Observer provided is both interesting and potentially troubling. It is interesting because, in searching for a notion of reference frame compatible with Shape Dynamics, we have naturally obtained a very simple holographic duality where both sides of the duality are completely understood. Moreover, straightforward generalizations are possible, such as adding interactions, considering a scalar field, or quantizing the theory. This would allow an interesting testing ground for gaining a deeper understanding of holographic dualities such as the AdS/CFT \cite{Maldacena:ads_cft,Witten} or the related dS/CFT \cite{Strominger:holo_cosmo,Skenderis:holo_uni} correspondences. On the other hand, the holographic properties of the description are potentially troubling because the duality between Shape Dynamics and General Relativity itself is expressed as a bulk-bulk duality, and not a bulk-boundary duality like the one presented here. The reliance in our formulation of Shape Observers on the properties of the conformal boundaries of dS is completely unnecessary in the construction of Shape Dynamics. Clearly, more work is necessary to see if this notion of Shape Observer can be made relevant for a local notion of inertial reference frame in Shape Dynamics. However, the results of this paper would seem to suggest that a strict equivalence between conformal frames in Shape Dynamics and reference frames in General Relativity may not exist because of a lack of representational equivalence between the relevant symmetry groups. If such an equivalence does exist, then it would seem that we are in need of some new insight to help make the connection. Without such an insight, one is led to conclude that the conformal invariance of Shape Dynamics should be thought of more as a ``complimentary'' or ``hidden'' symmetry of gravity rather than a ``dual'' symmetry to hypersurface deformations. A strict duality would only be seen to apply in a holographic setting.

Lastly, we can justify the drastic simplification we have made in considering the maximally symmetric case by noting that rigorous clock synchronization procedures are only really specifiable in the case where the metric is approximately Minkowski (the de~Sitter case we consider here is only slightly different in this regard). In this case, we can simply rely on clock synchronization procedures used for Special Relativity. Thus, the maximally symmetric case already has a lot of the important physics contained in it. Furthermore, Einstein--Cartan formulations of gravity \cite{Hehl:1994ue} and MacDowell--Mansuri-type approaches to General Relativity \cite{MacDowell:1977jt,Stelle:1979va,Wise:MM_paper,Randono:2010cq} show how the internal symmetry groups of the frame fields can be related, via the equations of motion and the soldering equation, to infinitesimal spacetime diffeomorphism invariance. Mathematically, this is perhaps because of the strong connection implied by Cartan geometry \cite{Sharpe:Cartan_geometry} between the internal symmetry groups of the frame fields and the local properties of the geometry (more on this is \S\ref{sec:SD and gen}). Thus, there are also mathematical reasons to argue for a strong connection between the internal symmetries of the frame fields, which are transformations relating local inertial observers, and the properties of a curved geometry.

\section{Prelimiaries: de~Sitter Spacetime}

To begin our presentation, we review the basics of dS spacetime and establish the notation we will use for the remainder of the paper, highlighting the features we will need for our construction of Shape Observers in \S~\ref{sec:holo SO}.

\subsection{Definition}

dS$^{D,1}$ spacetime is the maximally symmetric solution to the Einstein equation with positive cosmological constant. It is most easily represented by an embedding into $\mathbbm R^{D+1,1}$ where it can be treated as the hyperboloid satisfying
\begin{equation}\label{eq:ds def}
    g_{\mu\nu} x^\mu x^\nu = \ell^2,
\end{equation}
where $x^\mu$ are coordinates and $g_{\mu\nu}=\text{diag}(-1,1,\hdots,1)$ the flat metric on $\mathbbm R^{D+1,1,}$ (with $\mu,\nu = 0,\hdots D+1$). The dS radius $\ell$ is related to the cosmological constant, $\Lambda$, through $\Lambda = \frac {D(D-1)}{2\ell^2}$. Using the coordinates $x^I = (x^0,x^I)$ in $\mathbb R^{D+1,1}$, where $I = 1, \hdots, D+1$, we can choose a well-known embedding of the form
\begin{subequations}\label{eq:cmc slices}
\begin{align}
    x^0 &= \ell \sinh \alpha \\
    x^i &= \ell \cosh \alpha \, r \tilde x^i,  \\
    x^{d+1}  &= \ell \cosh \alpha \, w,
\end{align}
\end{subequations}
where $\alpha$ is a dimensionless hyperbolic angle, $w$ picks out a preferred direction in the ambient space, and $r$, $w$, and $\tilde x^i$ are chosen such that $r^2 + w^2 = 1$ and $\tilde x^i \tilde x^j \delta_{ij} = 1$. It follows from these definitions that $(r\tilde x^i, w)$ are coordinates on the unit $D$-sphere, and can be expressed in terms of the $D$ angles $\Omega_D$, while $\tilde x^i$ are coordinates on the unit $(D-1)$-sphere, and can be expressed in terms of the $D-1$ angles $\Omega_{D-1}$.

The preferred direction picked out by these coordinates will be understood as selecting the \emph{South Pole} at $w = -1$, $r = \tilde x^i = 0$ as the location of a stationary observer, $O$. Using the results of the Appendix, it is easy to see that $O$ is inertial (and corresponds to the curve traced out by $A^\mu = \frac \ell 2 (-1,\vec 0, -1)$ and $B^\mu = \frac \ell 2 (1,\vec 0, -1)$). This leads to a simple interpretation of these coordinates: they represent the coordinates attributed to events in dS spacetime in the rest frame of an inertial observer at the South Pole. What this observer sees are events occurring on a $D$-sphere that expands and contracts according to the hyperbolic cosine of $\alpha$, which, as can easily be shown, is proportional to the proper time, $\tau$, along the particles trajectory. Moreover, these coordinates are in one-to-one correspondence with the $x^\mu$'s that satisfy \eqref{eq:ds def}, and thus represent a global embedding of the spacetime.

The metric in terms of the coordinates \eqref{eq:cmc slices} is found by noting that the ambient metric
\begin{equation}
    ds^2 = -(dx^0)^2 + \delta_{IJ} dx^I dx^J
\end{equation}
leads to the induced metric
\begin{equation}\label{eq:cmc metric}
    ds^2 = \ell^2 \lf( -(d\alpha)^2 + \cosh^2\alpha\, d\Omega^2_D \rt),
\end{equation}
where $d\Omega_D$ is the line element for the $D$-sphere. The global Killing vectors of this metric generate all the transformations that leave \eqref{eq:ds def} invariant. These are, obviously, the $SO(D+1,1)$ Lorentz transformations in the ambient spacetime and can be used to obtain the trajectory of an arbitrary inertial observer by appropriately transforming the trajectory of the stationary observer.

The extrinsic curvature, $K_{IJ}$, of the constant-$\alpha$ $D$-spheres can be calculated from the metric \eqref{eq:cmc metric},
\begin{align}
	K_{IJ} &= \mathcal L_{n_\alpha}g_{IJ} = \frac {\ell^2\, \Omega_{IJ}^D} {\sqrt{-g_{00}}} \frac d {d\alpha} \cosh^2 \alpha \\
	       &=  2 \ell \cosh \alpha \sinh \alpha\, \Omega_{IJ}.
\end{align} 
The York time on these hypersurfaces, given by
\begin{equation}
	\tau_\text{York} = \frac D 2 g^{IJ} K_{IJ} = \frac 1 \ell \tanh \alpha,
\end{equation}
is constant and monotonic. Thus, the coordinates \eqref{eq:cmc slices} are closed, \emph{Constant Mean (extrinsic) Curvature}, or CMC, slices. Because the metric is invariant under ambient Lorentz transformations, all inertial observers related to $O$ will also be closed and CMC. The significance of this is that closed CMC slices are precisely the foliation condition, as shown in \cite{gryb:shape_dyn}, where General Relativity and Shape Dynamics are equivalent.

\begin{figure}
    \begin{center}
	\includegraphics[width= 0.75\textwidth]{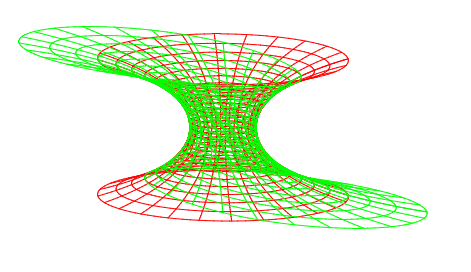}
    \end{center}
    \caption{$w$-boosts applied to the dS hyperboloid for $\psi = 0.5$. The $z$-axis shows proper-time.}
    \label{fig:bulk boosts}
\end{figure}

The CMC observer $O$ provides a way of visualizing how the Lorentz transformations act on the dS spacetime. By singling out this observer, we can identify the $SO(D+1)$ subgroup of $SO(D+1,1)$ that preservers the CMC hypersurfaces. These are the symmetries which act on $d\Omega_{D}$, the metric on the $D$-spheres. This $SO(D+1)$ subgroup, in turn, splits into the $SO(D)$ subgroup leaving $O$'s position fixed --- representing rotations around this observer --- and the $D$-dimensional translations that shift $O$'s the position to different places on the sphere --- representing translations of $O$. The boost subgroup is harder to visualize because it changes the definition of the constant-time hypersurfaces. For $O$ (and $O$ only), a $w$-boosts parameterized by the hyperbolic angle $\psi$, acts in the following way:
\begin{equation}
	x'^\mu(\alpha) = \Lambda^\mu_\nu x^\nu (\alpha) = \lf( \begin{array}{c c c} 
				\cosh \psi & 0 & -\sinh \psi \\
				0 & 0 & 0 \\
				-\sinh \psi & 0 & \cosh \psi
			    \end{array}\rt)
			    \lf( \begin{array}{c}
				\ell \sinh \alpha \\ 0 \\ \ell \cosh \alpha
			    \end{array}\rt) = \lf( \begin{array}{c}
				    \ell \sinh(\alpha - \psi) \\ 0 \\ \ell \cosh(\alpha - \psi).
			    \end{array}\rt),
\end{equation}
which is a time translation $\alpha \to \alpha - \psi$ along $O$'s trajectory. The $D$ remaining boosts in the ambient spacetime can be interpreted as ``boosts'' in dS spacetime. However, it is important to keep in mind that these ambient boosts only act on dS like ``time translations'' and ``boosts'' for $O$. For other observers, their action is more complicated. Figure~\ref{fig:bulk boosts} shows how a $w$-boost acts on all points of the dS hyperboloid. The effect of the other boosts can be obtained by applying translations on the $D$-spheres.

To close this section, we note that dS spacetime has a particle and an event horizon due to the presence of spacelike conformal boundaries in the past and future (which are reached in infinite proper time). These can be seen by going to conformal time, $T$,
\begin{equation}
	\sec\lf( \frac T \ell \rt) = \cosh \alpha,
\end{equation}
where the metric takes the form
\begin{equation}
	ds^2 = \frac 1 {\cos^2 \frac T \ell} \lf( -dT^2 + \ell^2 d\Omega^2_D \rt)
\end{equation}
and $-\frac \pi 2 \leq \frac T \ell \leq \frac \pi 2$. Since conformal transformations do not affect null geodesics, we see that $ \pi + 2T$ and $\pi - 2T$ measure, respectively, the size of the particle and event horizons at any point in the spacetime. The existence of these horizons will lead to the shape freezing described in \S\ref{sec:Shape Freezing}, which will be key to our construction. Since the range of $T$ is only half that of the range of angles on the $\De$-sphere, a null geodesic can travel at most across half of space, implying that the past horizon can only cover the whole of space on the future conformal boundary. 

\subsection{Inertial Observers}\label{sec:Inertial Observers}

The trajectories of inertial observers in dS are given by timelike geodesics. In Appendix~\ref{apx:geodesics}, we compute the trajectories of arbitrary timelike geodesics in dS whose trajectories are given by extremizing the action
\begin{equation}
	S = \int_{x^\mu(t_1)}^{x^\mu(t_2)}  \de t \lf[ {\textstyle \frac 1 2} m\sqrt{ - \eta_{\mu\nu} \dot x^\mu \dot x^\nu } + \lambda \lf( \eta_{\mu\nu} x^\mu x^\nu - \ell^2 \rt) \rt]\,,
\end{equation}
which is taken to be the proper time along a geodesic between two points $(x^\mu(t_1), x^\mu(t_2))$ for some point particle of mass $m$. All coordinates are conveniently defined in the ambient $\mathbb R^{(\De + 1,1)}$ spacetime. The Lagrange multiplier $\lambda$ enforces the constraint that these coordinates are restricted to the dS hyperboloid in this ambient spacetime.

\begin{figure}
	\centering
	\begin{subfigure}{.95\linewidth}
		\centering
		\includegraphics[width = \textwidth]{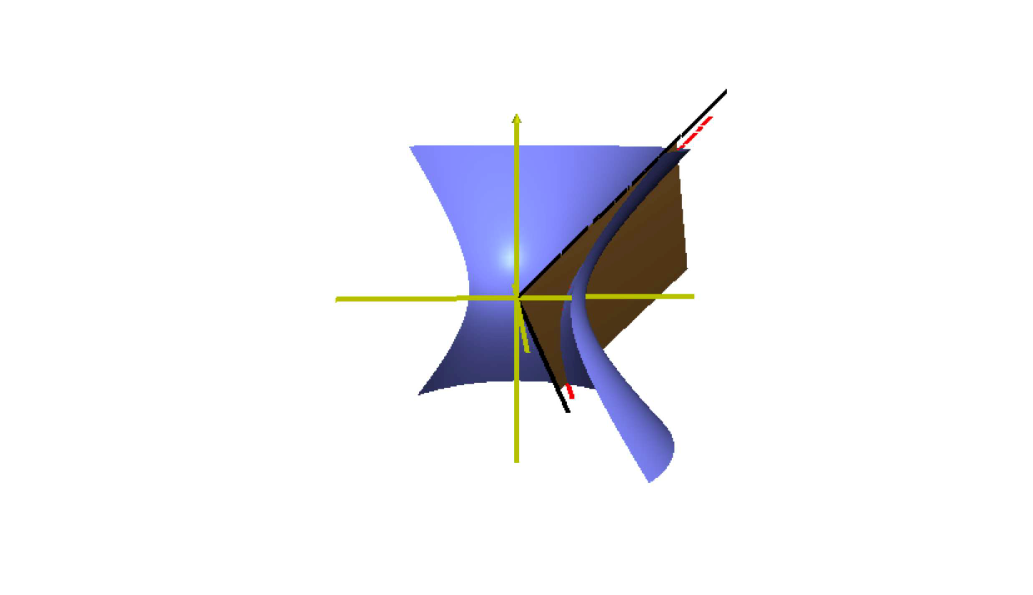}
		%\label{}
	\end{subfigure}
	\begin{subfigure}{.95\linewidth}
		\centering
		\includegraphics[width = \textwidth]{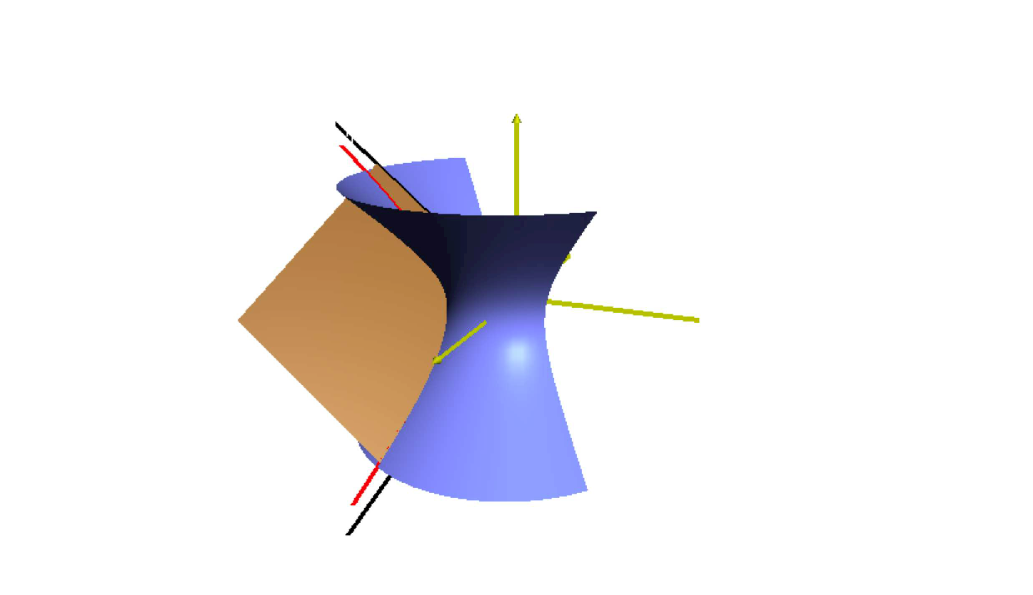}
		%\label{}
	\end{subfigure}	
	\caption{Two views of an inertial observer's worldline in dS. The worldline, in red, is determined by the intersection of dS, in blue, with the span, in brown, of two backward and forward pointing null vectors, in black.} \label{fig:geodesics}
\end{figure}

The solutions can be parametrized in terms of two null vectors --- one `backward pointing', $\xi\in^\mu$, and one `forward pointing', $\xi\out^\mu$ --- that obey the normalization condition
\begin{equation}
	2\eta_{\mu\nu} \xi\in^\mu \xi\out^\mu = \ell^2\,.
\end{equation}
In terms of these, the general solution is
\begin{equation}\label{eq:gen soln}
	x^\mu(\tau) = \xi\out^\mu e^{\sfrac m \ell \tau } + \xi\in^\mu e^{-\sfrac m \ell \tau }\,,
\end{equation}
using a proper-time parametrization, $\tau$. A convenient and intuitive way to understand these trajectories is to note that they are given by the intersection of the 2-plane going through the origin that is spanned by $\xi\in^\mu$ and $\xi\out^\mu$ and the dS hyperboloid in the ambient spacetime, in analogy to the great circles of a sphere. It is a relatively straightforward exercise to check that the solution \eqref{eq:gen soln} is indeed tracing out the intersection this 2-plane with the dS hyperboloid. Figure~\ref{fig:geodesics} shows an example of how a typical timelike geodesic can be constructed.

\subsection{Euclidean De-Compactification}\label{sec:inst shapes}

In order to establish a notion of Shape Observers, we will require a de-compactification of the CMC $D$-spheres onto the Euclidean plane. This will both provide us with the ability to define a traditional notion of ``shape'' for configurations of particles in the system and allow us to study the action of the dS symmetries in terms of the action of the conformal group on the plane. The particular de-compactification we will use is given by the coordinates, $X^i$, defined by
\begin{equation}\label{eq:de-compact}
	X^i = \frac{x^i}{x^0 - x^{D+1}}.
\end{equation}

These coordinates conveniently give us a bulk expression that reduce to a standard stereographic projection of the D-sphere onto a plane on the boundary. To see that this is indeed the case, note that (as is shown more explicitly in \S~\ref{sec:holo SO}) the dS hyperboloid becomes nearly light-like in the ambient spacetime near the conformal boundary. In this limit, the size of the projected sphere drops out of \eqref{eq:de-compact} because $|\bar x^I| \to |\bar x^0|$ due to the null condition. Thus,
\begin{subequations}\label{eq:asym X}
\begin{align}
	X^i &\to \frac{\bar{\tilde x}^i}{1 - \bar w}, \quad \text{for } x^0 > 0 \\
	    &\to -\frac{\bar{\tilde x}^i}{1 + \bar w}, \quad \text{for } x^0 < 0.
\end{align}
\end{subequations}
In the $x^0 < 0$ case, the above equation represents a standard stereographic projection taken with respect to the North Pole onto a plane passing through the equator. Conversely, in the $x^0 > 0$ case, the above equation represents a stereographic projection of the anti-podal point to $\bar x^I$.\footnote{There is a problem with these coordinates at $x^0 - x^{D+1} = 0$, but this will not be a concern for us because we will mainly be interested in the dS boundaries where $|x^0 - x^{D+1}| \gg \ell$.}

This particular de-compactification will be interesting for our purposes because of its transformation properties under infinitesimal Lorentz transformations. We represent these by
\begin{equation}\label{eq:X new def}
	x^\mu \to x^\mu + \omega^\mu_\nu x^\nu,
\end{equation}
where $\omega_{\mu\nu} = -\omega_{\nu\mu}$. Using the CMC observers described in the last section, we can interpret these symmetries by how they act on the stationary observer $O$: $\w i j$ represent $SO(D)$ rotations of $O$ on the CMC surfaces, $\w i {D+1}$ represent spatial translations of $O$ on the sphere, $\w 0 {D+1}$ represents time translations of $O$ along its trajectory, and $\w i 0$ represent ``boosts''. In terms of the ambient coordinates, these transformations split up as:
\begin{subequations}
\begin{align}
	x^0 &\to x^0 + \w 0 j x^j + \w 0 {D+1} x^{D+1} \\
	x^i & \to x^i + \w i j x^j + \w i 0 x^0 + \w i {D+1} x^{D+1} \\
	x^{D+1} & \to x^{D+1} + \w {D+1} j x^j + \w {D+1} 0 x^0.
\end{align}
\end{subequations}

The decompactified coordinates are conveniently expressed in terms of the light cone coordinates
\begin{equation}
	x^\pm = x^0 \pm x^{D+1},
\end{equation}
so that $X^i = \frac {x^i} {x^-}$.\footnote{There is a dual set of coordinates $X^i = \frac {x^i} {x^+}$, where the alternative light cone coordinate $x^+$ transforms with the opposite weight under dilations. The choice of $x^-$ versus $x^+$ is purely conventional.} Under Lorentz transformations \eqref{eq:X new def}, $x^-$ transforms as
\begin{equation}
	x^- \to x^- - \w {D+1} 0 x^- + \lf( \w 0 j - \w {D+1} j \rt) x^j.
\end{equation}
while $X^i$ transforms as
\begin{equation}
	X^i \to X^i + \w i j X^j + \w {D+1} 0 X^i + \frac{\w i 0 x^0 + \w i {D+1} x^{D+1}}{x^-} + \lf( \w {D+1} j - \w 0 j \rt) X^i X^j
\end{equation}
(where we used $\w {D+1} 0 = \w 0 {D+1}$). The above transformations suggest the following definitions
\begin{align}\label{eq:isomorphism}
	d &\equiv \w 0 {D+1} = \w {D+1} 0 & t^i &\equiv \frac 1 2 \lf( \w i 0 - \w i {D+1} \rt) & s^i \equiv \frac 1 2 \lf( \w i 0 + \w i {D+1} \rt).
\end{align}
These imply
\begin{align}
	t_i &= \frac 1 2 \lf( \w 0 i + \w {D+1} i \rt)  & s_i &= \frac 1 2 \lf( \w 0 i - \w {D+1} i \rt) 
\end{align}
and
\begin{align}
	\w i 0 &= \lf( s^i + t^i \rt) & \w i {D+1} &= \lf( s^i - t^i \rt)\,.
\end{align}
In terms of these new quantities, we find
\begin{align}
	x^- &\to x^- - d x^- + 2s_i x^j \label{eq:x- transf} \\
	X^i &\to X^i + \w i j X^j + d X^i + t^i + s_j \lf( \delta^{ij} \frac{(x^0)^2 - (x^{D+1})^2}{(x^-)^2}  - 2X^i X^j\rt). \label{eq:conf trans X}
\end{align}

The transformations above have a straightforward interpretation in terms of their action on the decompactified coordinates $X^i$. The $\w i j$ represent rotations and $t^i$ translations of $X^i$ on the plane. The $d$'s represent dilations of $X^i$ while the $s^i$ are still somewhat obscure. It is straightforward to see that these will parametrize special conformal transformations near the dS boundary where the ambient coordinates become null. However, in the bulk they don't admit such an interpretation.

\subsection{An Obstacle for Bulk ``Shape Observers''}

Given the interpretation of \eqref{eq:conf trans X} in terms of conformal symmetries, it is tempting to try to use the $X^i$'s to define a notion of bulk observer who only see shape information --- expressed in terms of the $X^i$ coordinates of some system of particles. However, there are several reasons why this notion is inadequate.

Firstly, although the $X^i$'s transform in the bulk under the similarity group (which is the Galilean group plus dilatations), it also transforms under the unusual symmetry parametrized by the $s^i$'s, which has no direct interpretation in terms of conformal symmetries. Simply disregarding the symmetry is also not natural because it would restrict to only certain inertial observers in dS spacetime.

Secondly, and more problematically, a shape must be defined at a particular instant. This could be represented by a configuration of point particles with one set of $X^i$ coordinates per particle at some time. However, there is no choice of time that would be invariant under all the dS symmetries. This means that the time coordinates for different particles in the configuration will transform differently under the bulk dS isometries. The $x^-$ coordinate, for example --- which is perhaps a natural choice from the conformal point of view --- transforms differently for different particles because of the last term in \eqref{eq:conf trans X} (again, the special conformal transformations are the problem).\footnote{There is also the problem that the spacetime metric is singular at $x^- = 0$.} What this means is that different bulk inertial observers don't agree on what constitutes an instantaneous configuration, because they don't share the same notion of simultaneity.

The difficulty we are encountering is that, although there is a formal isomorphism between the isometry group of dS$^{D,1}$ spacetime and the conformal isometries of the Euclidean plane $\mathbbm R^D$, the \emph{action} of these groups on point particles in dS does not, in general, translate into the \emph{action} of the conformal group on the $X^i$'s. Perhaps this could be cured by choosing a different set of bulk coordinates?\footnote{In fact, the coordinate choice $\bar X^i = \frac {x^i}{\frac {x^0}\gamma - x^{D+1} }$, where $\gamma = \sqrt \frac{(x^0)^2}{(x^0)^2 + \ell^2 } $, does have the right transformation properties in bulk. This can be seen by noting that these coordinates represent a projection of the dS coordinates onto the light-cone, which is conformally invariant, in the ambient spacetime along surfaces of constant $x^0$.} However, the dS symmetries also act on the time coordinates of the particles, and this action cannot be represented in terms of the action of the conformal group. This leads to a fundamental representational inequivalence between the bulk dS isometries and the conformal group in the bulk. Fortunately, near the conformal dS boundaries, the spacetime becomes indistinguishable from the ambient light cone in a way we will state precisely below. In this region, the metric becomes degenerate in the time direction defined by $x^-$, leading to a phenomenon we call ``shape freezing''. Thus, in these regions, the dS isometries do act conformally on the configurations $X^i$. In the next section, we will spell this out explicitly and use these facts to define a notion of Shape Observer.

\section{Holographic Shape Observers}\label{sec:holo SO}

The asymptotic regions of dS are defined, in the ambient coordinates, by the condition
\begin{equation}\label{eq:asymptotic}
	\frac{(x^0)^2} {\ell^2} \gg 1.
\end{equation}
This defines two separate regions: the future conformal boundary, represented in these coordinates by $+x^0/\ell \gg 1$ and past conformal boundary, $-x^0/\ell \ll -1$. In the following, we will distinguish these two regions using a $\pm$ or $\mp$ notation, and adopt the convention (where appropriate) that the top symbol refers to the future while the bottom symbol to the past.

In the limit \eqref{eq:asymptotic}, the ambient coordinates $x^\mu$ become approximately null
\begin{equation}\label{eq:null}
	x^\mu x_\mu \approx 0
\end{equation}
as the dS hyperboloid becomes indistinguishable from the ambient light cone. Furthermore, the spacetime metric becomes degenerate. This is most easily seen by first performing the change of coordinates
\begin{equation}
	x^\mu \to \lf( x^+, x^-, x^i \rt),
\end{equation}
where the metric takes the form
\begin{equation}\label{eq:pm metric}
	g = \lf( 
		\begin{array}{c c c}
			0 & -\frac 1 2 & 0 \\
			-\frac 1 2 & 0 & 0 \\
			0 & 0 & \delta_{ij}
		\end{array}
	\rt).
\end{equation}
Then, the null condition
\begin{equation}
	g_{\mu\nu} x^\mu x^\nu = - x^+ x^- + x^i x_i \approx 0
\end{equation}
can be used to solve for $x^+$ in terms of the other components
\begin{equation}
	x^+ = \frac{x^i x_i} {x^-}.
\end{equation}
We can use this to express the ambient coordinates $x^\mu$ in terms of the decompactified coordinates $X^i$ and the light-cone coordinate $x^-$ as
\begin{align}\label{eq:X to x}
	x^i &= x^- X^i & x^0 &= x^- ( X^i X_i + 1 ) & x^{D+1} &= x^- ( X^i X_i - 1 )\,.
\end{align}
In terms of these coordinates,
\begin{equation}
	g_{\mu\nu} dx^\mu dx^\nu = (x^-)^2 (dX^i)^2\,.
\end{equation}
All terms containing $dx^-$ drop out, implying that the metric is degenerate in this direction. This approximate degeneracy leads to important physical consequences that we will explore in the next subsection.

\subsection{Shape Freezing}\label{sec:Shape Freezing}

The phenomenon that we will call \emph{shape freezing} will refer to a property of the dynamics of particles in dS spacetime whereby the expansion of space is such that massive objects are effectively `outrun' by space in such a way that the shape created by configurations of particles on the $\De$-sphere becomes effectively \emph{frozen}. Intuitively, this is because, at the conformal boundary, the dS hyperboloid becomes closer and closer to the light-cone in the embedding spacetime so that space is basically expanding at the `speed of light'. We can make these statements precise by breaking the argument into three steps: i) first show that all choices inertial observers are effectively equivalent in this region, or that \emph{foliation freezing} occurs, ii) then show how the configurations of a system of massive particles gets frozen after decompactification, or that \emph{configuration freezing} occurs iii) and, finally, show that these frozen configuration are indeed invariant under conformal transformations, so that the frozen configurations represent frozen shapes.

\subsubsection{Foliation Freezing}

Our first task is to show that the choice of inertial observer becomes irrelevant near the conformal boundaries. This means that ambient boosts will no longer appreciably change the notion of simultaneity of different boosted observers. To make this more precise, consider the \emph{finite} sized boost parametrized by $\psi$. Since $\psi$ is finite, there always exists a time coordinate $\alpha$ such that $\alpha \gg \psi$. Since, using \eqref{eq:cmc slices}, we have
\begin{equation}
	\cosh \alpha = \sqrt{1 + \lf(\frac{x^0}{\ell}\rt)^2} \approx \pm \frac{x^0}{\ell} = \pm \sinh \alpha.
\end{equation}
Furthermore, in the future boundary, $e^\alpha \gg 1$ while, in the past boundary, $e^{-\alpha} \gg 1$. Thus,
\begin{equation}
	\cosh \alpha \approx \pm \sinh \alpha \approx \frac 1 2 e^{\pm\alpha}.
\end{equation}
Without, loss of generality, we can consider a boost in the $w$-direction (since all other boosts can be obtained through a translation on the $\De$-sphere). Under such a boost, $x^0$ transforms as
\begin{align}
	\sinh \alpha = \frac {x^0} \ell &\to \frac {x^0} \ell \cosh \psi + \frac {x^{D+1}} \ell \sinh \psi \\
		&= \sinh \alpha \cosh \psi + w \cosh \alpha \sinh\psi.
\end{align}
A few simple rearrangements lead to
\begin{equation}
	\alpha \to \alpha + \ln \lf( \cosh \psi \pm w \sinh \psi \rt).
\end{equation}
The linear term always dominates over the logarithm as can be seen by taking the limit $\alpha \gg \psi \gg 1$ and $w=1$ (the largest value $w$ can have). Then, we can similarly use
\begin{equation}
	\cosh \psi \approx \pm \sinh \psi \approx e^{\pm \psi}
\end{equation}
so that
\begin{equation}
	\alpha \to \alpha \pm 2\psi \approx \alpha.
\end{equation}
This means that one can always wait a sufficiently long time so that, $\alpha$, will always dominate over any \emph{finite} boost $\psi$, implying that all physical inertial observers will effectively see the same foliation at late times.

\subsubsection {Configuration Freezing}

In \S\ref{sec:inst shapes} we saw explicitly from equations \eqref{eq:asym X} that
\begin{equation}\label{eq:asymp X}
	X^i \approx \frac{ r\tilde x^i} {\pm 1 - w}\,,
\end{equation}
which is just the stereographic projection of the $\De$-sphere near the conformal boundaries onto a plane. From this expression, it is clear that, asymptotically, the $X^i$'s loose their $x^0$-dependence, leading to a freezing out of configurations.

\subsubsection {Shape Freezing}

Using the bulk transformations properties of the $X^i$ under the ambient Lorentz transformations derived in \S\ref{sec:inst shapes}, we can now take the asymptotic limit of these transformations. The last term of equation~(\ref{eq:conf trans X}) (which is the bulk transformation equation for the $X^i$'s under the dS isometries) can be re-written using the null condition \eqref{eq:null}, giving us
\begin{equation}
	x^I x_I = (x^0)^2 - (x^{D+1})^2.
\end{equation}
Then, the $s^i$ term in equation~(\ref{eq:conf trans X}), becomes
\begin{equation}
	\delta^{ij} X^k X_k  - 2X^i X^j,
\end{equation}
which is now the genuine generator of special conformal transformations. Thus, in this limit, the bulk $SO(D+1,1)$ transformations act like genuine conformal transformations.

If we now consider a system of many particles with coordinates $x^\mu_\Upsilon$, where $\Upsilon = 1,\hdots, n$ labels the different particles, then the collection of such coordinates at some instant of time $t$, defined by some timelike hypersurface, forms an instantaneous configuration of the system. Near the conformal boundaries, since all timelike hypersurfaces formed by the constant proper-time hypersurfaces of a congruence of inertial observers will effectively define the same notion of time, the instantaneous configurations can simply be given by the decompactified $X^i_\Upsilon$'s, which are time-independent. Because the $X^i_\Upsilon$'s are also conformally invariant in these regions, the gauge-independent information given by these configurations is simply given by the scale-invariant \emph{shape} of the system. This leads directly to the notion of \emph{shape freezing} we were looking for.

\begin{figure}
	\centering
	\begin{subfigure}{.45\textwidth}
		\centering
		\includegraphics[width = \textwidth]{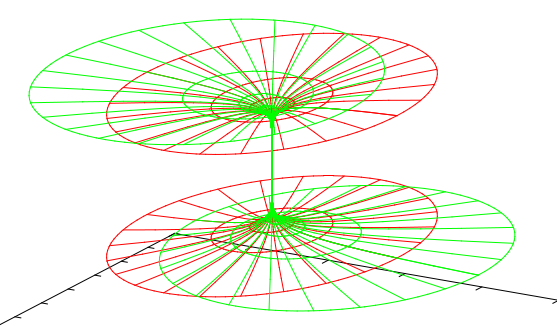}
		\caption{Boosts will stretch the $x^i$'s leading to dilatations of the $X^i$'s.}
		%\label{}
	\end{subfigure}
	\begin{subfigure}{.45\textwidth}
		\centering
		\includegraphics[width = \textwidth]{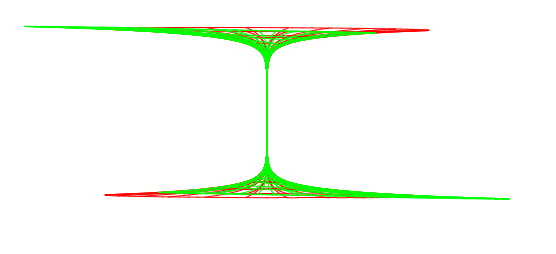}
		\caption{Finite boosts don't change appreciably the notion of simultaneity near the boundary.}
		%\label{}
	\end{subfigure}	
	\caption{The asymptotic ``freezing-out'' of shapes in dS. The green and red surfaces represent a region of dS spacetime before and after boosting.} \label{fig:freezeout}
\end{figure}

This shape freezing is a direct result of the isomorphism between $SO(D+1,1)$ and $Conf(D)$, but now the \emph{action} of the ambient Lorentz transformations on the coordinates $x^i$ can be mapped to the \emph{action} of the conformal group on the decompactified coordinates $X^i$. This would not be possible without also having the notion of foliation freezing, so that the $X^i$'s can be treated as having roughly the same time parameter. What we find is that the ambient Lorentz transformations are important enough to appreciably change the \emph{spatial} scale of the instantaneous configurations without appreciably changing the notion of an instant. In Figure~\ref{fig:freezeout}, these effects are illustrated for a finite boost $\psi = 0.1\alpha$. The spatial profiles near the boundary are noticeably effected by the boost --- in a way that leads to a representational isomorphism with the conformal group --- while the instantaneous hypersurfaces are hardly effected.

\subsection{A Holographic Shape Dynamics Theory}

Our strategy for defining a theory of dynamical conformally invariant shapes using the physics of bulk inertial observers in dS spacetime will be to construct the Hamilton--Jacobi function for the conformal theory by using the asymptotic properties of the bulk action evaluated along a particular solution. Specifically, we will take the bulk action for massive particles in dS spacetime and evaluate this when the trajectories start and end near the conformal boundaries. Because of the shape freezing we discussed in the previous sections, this will effectively give a function that depends on the initial, \emph{``in''}, and final, \emph{``out''}, shapes of the system. As we will see, this function has all the same properties as the Hamilton--Jacobi function for a conformally invariant theory.

To be precise, we define the Hamilton--Jacobi function, $S$, by summing over the proper time along a geodesic (scaled appropriately by the mass) for an $N$-particle system, using a particular notion of simultaneity, and then take the limit where the time function, $t$, on our surfaces of simultaneity is large compared to the dS horizon scale, $\ell$. We then define the decompactified coordinates $X^{i\Upsilon}$ on the surfaces of simultaneity, where $\Upsilon = 1 \hdots N$ is a particle label, so that the dS isometries act like the conformal group. This leads formally to
\begin{equation}\label{eq:holo defn}
	S(X^{i\Upsilon}\in,X^{i \Upsilon}\out ) = \lim_{\frac{t_0}{\ell} \to \infty}\, \sum_{\Upsilon = 1}^N\int_{x^\mu(-t_0, X\in)}^{x^\mu(t_0, X\out)} dt_\Upsilon\, \sfrac m 2 \sqrt{g_{\mu\nu} \dot x^\mu_\Upsilon \dot x^\nu_\Upsilon}\,,
\end{equation}
where $X^{i\Upsilon}\outin$ represent the ``in'' and ``out'' shapes of the system. It will be convenient for us to use as a time coordinate the light-cone coordinate $t = x^-$, so that the $\frac {|t|} \ell \to \infty$ limit corresponds to $|x^-| \gg \ell$.

In an effort to reduce the notation, we will first consider the single particle case before easily generalizing to the many particle case. Recall from the discussion in \S\ref{sec:Inertial Observers} and from the results derived in Appendix~\ref{apx:geodesics} that a proper time parametrization of a geodesic observer in dS has coordinates and velocity given by
\begin{align}
	x^\mu(\tau) &= \xi^\mu\in e^{\sfrac m \ell \tau } + \xi^\mu\out e^{-\sfrac m \ell \tau }\\
	\dot x^\mu(\tau) &= \frac m \ell \lf[ \xi\in^\mu e^{\sfrac m \ell \tau } - \xi\out^\mu e^{-\sfrac m \ell \tau } \rt]\,.
\end{align}
Inserting this into the action \eqref{eq:holo defn} (restricted to a single particle) gives straightforwardly
\begin{equation}
	S = \frac {m^2} 2 \Delta \tau\,,
\end{equation}
highlighting that $\tau$ is just proportional to the proper time.

It remains to compute the difference in proper time between two arbitrary points in the bulk $(x_1^\mu(\tau_1), x_2^\mu(\tau_2))$ lying along the same geodesic defined by the null vectors $(\xi^\mu\in,\xi^\mu\out)$
\begin{align}
	x_1^\mu(\tau_1) &= \xi\in^\mu e^{\sfrac m \ell \tau_1} + \xi\out^\mu e^{-\sfrac m \ell \tau_1} \\
	x_2^\mu(\tau_2) &= \xi\in^\mu e^{\sfrac m \ell \tau_2} + \xi\out^\mu e^{-\sfrac m \ell \tau_2}.
\end{align}
Recalling from \S\ref{sec:Inertial Observers} that $2\xi\in\cdot \xi\out = \ell^2$ (where we have used the notation $A \cdot B = \eta_{\mu\nu} A^\mu B^\nu $) and using the fact that these vectors are null, we find
\begin{equation}
	\frac{x_1\cdot x_2}{\ell^2} = \cosh \sfrac m \ell(\tau_1 - \tau_2)\,.
\end{equation}
Thus,
\begin{equation}
	\Delta \tau = \sfrac \ell m \cosh^{-1} \lf( \frac {x_1 \cdot x_2} {\ell^2} \rt)
\end{equation}
or
\begin{equation}\label{eq:exact S}
	S = \sfrac {m\ell} 2 \cosh^{-1} \lf( \frac {x_1 \cdot x_2} {\ell^2} \rt)\,.
\end{equation}

The next step is to express $x_1\cdot x_2$ in terms of the light-cone coordinate $x^-$ and the decompactified coordinates $X^i$ before taking the asymptotic limit. To do this, we use the metric \eqref{eq:pm metric}, to obtain
\begin{equation}
	x_1\cdot x_2 = -\sfrac 1 2 \lf( x_1^+ x_2^- + x_1^- x_2^+  \rt) + \vec x_1 \cdot \vec x_2\,,
\end{equation}
where $\vec x_1 \cdot \vec x_2 = \delta_{ij}\, x_1^i x_2^j$. We can use the dS condition $x\cdot x = \ell^2$ to solve for the other light-cone coordinate $x^+$
\begin{equation}
	x^+ = \frac {\vec{x}^2 - \ell^2} {x^-} = \frac 1 {x^-} \lf( \frac{X^2}{(x^-)^2} - \ell^2 \rt) \,.
\end{equation}
Using this and $x^i = X^i x^-$, we find
\begin{equation}
	x_1\cdot x_2 = \frac 1 2 \lf[ \frac {\ell^2 \lf( (x^-_2)^2 + (x^-_1)^2 \rt)}{x^-_1 x^-_2} - x_1^- x_2^- \lf( X_1 - X_2 \rt)^2 \rt]\,.
\end{equation}
If we take the limit in such a way that our future and past boundaries a defined symmetrically about $x^- = 0$; i.e., so that
\begin{equation}
	x_2^- = - x_1^- \equiv t;
\end{equation}
then
\begin{equation}
	\boxed {S = \sfrac {m\ell} 2 \cosh^{-1} \lf[ \sfrac 1 2 \lf( \sfrac t \ell \rt)^2 (X_1 - X_2)^2 - 1 \rt] }\,.
\end{equation}

This integral is divergent as $(t/\ell) \gg 1$, or as $\epsilon = (\ell/t) \to 0$. However, we can understand the leading order behaviour of this divergence by making use of the well-known Laurent series expansion
\begin{align}
	\cosh^{-1} x &= \ln 2x - \sum_{n=1}^\infty \lf( \frac{(2n)!}{2^{2n} (n!)^2}\rt) \frac{x^{-2n}}{2n} \notag \\
				 &= \ln 2x - \lf( \sfrac 1 4 x^{-2} + \sfrac 3 {32} x^{-4} + \hdots \rt)\,.\label{eq:acosh expansion}
\end{align}
This leads to the following asymptotic expansion for $S$
\begin{equation}
	S = \sfrac {m\ell} 2 \lf[ \ln\lf( \frac{(X_1 - X_2)^2}{\epsilon^2} - 2 \rt) - \frac {\epsilon^4}{(X_1 - X_2)^4} + \hdots \rt]\,.
\end{equation}
Taking the $\epsilon \to 0$ limit of the above expression allows us to identify $X^i_1 \to X^i\in$ and $X^i_2 \to X^i\out$. If we then sum over many particles, we obtain
\begin{equation}
	\boxed{S(X^{i\Upsilon}\in, X^{i\Upsilon}\out) = M\ell \lf[ - \ln \epsilon + \ln \lf( \prod_\Upsilon \lf| \vec {X}^\Upsilon\in - \vec{X}^\Upsilon\out \rt|^{1/N} \rt) \rt]}\,,
\end{equation}
where $\epsilon = \prod_\Upsilon {\epsilon_\Upsilon}^{1/N}$ is a global time parameter for the system and $M = Nm$ is its total mass. It can simply be used as a global time label provided one uses the relational transformation properties under conformal transformations outlined in \S\ref{sec:Conformal Invariance}. The expression above is the main expression we are looking for. Note the logarithmic divergence of $S$ in proper time characterized by the pole $-M\ell \ln \epsilon $, which will be important for our considerations in \S\ref{sec:Conformal Invariance}.

We can further analyse the behaviour of the boundary theory by taking the lowest order terms of the Hamilton--Jacobi equations of motion for this system, in a case of a single particle (for simplicity). This is trivial because the Hamilton--Jacobi function, $S$, has already been computed. We find
\begin{align}
	\vec P = \diby S {\vec X_2} &\approx m\ell \frac{\lf( \vec X_2 - \vec X_1 \rt)}{(X_2 - X_1)^2} \lf[ \frac 1 {1-\frac{2 \epsilon^2}{(X_2 - X_1)^2}} \rt] + \hdots\, \notag\\
	  &\approx m\ell \frac{\lf( \vec X_2 - \vec X_1 \rt)}{(X_2 - X_1)^2}\,.\label{eq:HJ momenta}
\end{align}
Note that, to leading order, the $\epsilon$ dependence drops out of the equations of motion although it persists in $S$. This is the signature that we have a reparametrization invariant theory on the boundary (see the \S\ref{sec:Conformal Invariance} for a more substantive discussion on this point). Furthermore, the momentum constants of motion $\vec P$ obey identities in this limit:
\begin{equation}\label{eq:quad id}
	P^2 - \frac {m^2\ell^2}{ \Delta X^2} = 0.
\end{equation}
This has the same form as the Hamilton--Jacobi equation for a scale-invariant particle which has a Hamiltonian constraint of the form
\begin{equation}
	p^2 - \frac {m^2\ell^2} {q^2} = 0.
\end{equation}
These identities mean that $P^2$ cannot be determined uniquely in terms of the initial data for the theory. This is a further indication of a reparametrization invariant theory, where the indeterminacy is associated with the freedom to choose an arbitrary reference clock with which to define velocities for the system.

\subsection{Conformal Invariance}\label{sec:Conformal Invariance}

As will be shown in detail in the coming \S\ref{sec:Conformal Invariance}, the logarithmic divergence $m\ell \ln \epsilon$ is crucial for the conformal invariance of the classical theory. Thus, any renormalization of the boundary theory, which would subtract off such a divergence, will necessarily break conformal invariance. Thus, 

The boundary expansion of the action \eqref{eq:exact S} can be expressed in a Lorentz invariant way by noting that the inner product $\frac{x\in \cdot x\out}{\ell^2}\gg 1$ near the boundary. If we are only interested in the leading order contribution, then we have from \eqref{eq:exact S} and \eqref{eq:acosh expansion}
\begin{equation}
	S \approx \sfrac {m\ell}2 \ln \lf( 2 \frac {x\in \cdot x\out} {\ell^2} \rt)\,,
\end{equation}
where we are restricting to the single particle case again for simplicity of notation. In this limit, we can express the inner product $x\in \cdot x\out$ using the fact that $x\in^\mu$ and $x\out^\mu$ become null. In this limit, we have again that $x^+ = x^- X^2$ and that
\begin{align}
	x\in\cdot x\out &= -\sfrac 1 2 \lf( x\in^+ x\out^- + x\in^- x\out^+ \rt) + \vec x\in \cdot \vec x\out \notag \\
	&= -\frac {x\in^- x\out^-} 2 \lf( X\in - X\in \rt)^2\,.
\end{align}
Thus,
\begin{equation}\label{eq:boundary S}
	\boxed{S \approx \sfrac {m\ell}2 \ln \lf[ - \frac {x\in^- x\out^-} {\ell^2} \lf( X\in - X\out \rt)^2 \rt]}\,,
\end{equation}
which is valid only near the boundary. This expression allows us to consider the case where $x_1^- \neq -x_2^-$, but still very large.

It is now a straightforward exercise to check the conformal invariance of \eqref{eq:boundary S}. Since $S$ is a function only of the particle separations $|\vec X\in - \vec X\out |$, we immediately have that it is invariant under the Euclidean subgroup of the conformal group. The invariance of $S$ under dilatations and special conformal transformations requires the transformation properties of $x^-$ under the dS isometries. These can easily be worked out from the transformations \eqref{eq:x- transf} and the relations \eqref{eq:X to x} and are given by
\begin{equation}\label{eq:x- conf}
	x^- \to x^- \lf( 1 - d + 2 s_i X^i \rt),
\end{equation}
where $d$ is the dilatation parameter and $s_i$ is the infinitesimal parameter for special conformal transformations.

For dilatations, we can exponentiate \eqref{eq:x- conf} (with $s_i = 0$) to find that $x^- \to e^{-d} x^-$. Similarly, exponentiation of \eqref{eq:conf trans X} (with only $d$ non-zero) gives $X^i \to e^d X^i$. From this, it is clear that \eqref{eq:boundary S} is invariant.

For special conformal transformations, it is easiest to work with infinitesimal transformations since the formal exponentiation of \eqref{eq:x- conf} is non-trivial for $s_i \neq 0$. If we set $d = 0$ and use
\begin{equation}
	X^i \to X^i + s_j \lf( \delta^{ij} X^2 - 2X^i X^j \rt),
\end{equation}
then it is a short calculation to verify that \eqref{eq:boundary S} is invariant. Thus, we have explicitly verified that $S$ is conformally invariant near the boundary, as expected.

It is clear from the above derivation that the conformal invariance of $S$ depends importantly on the transformation properties of the time function $x^-$. One might then be curious about the role of $x^-$ in the boundary theory. However, we can see from the form of \eqref{eq:boundary S} that the logarithm allows us to write the Hamilton--Jacobi function of the boundary theory in the form
\begin{equation}
	S = f(x^-\in, x^-\out) + g(X\in,X\out)\,.
\end{equation}
Because $x^-$ plays the role of time in the boundary theory, this takes the form of a Hamilton--Jacobi function of a theory with a Hamiltonian that is time independent.

Indeed, the conformally invariant boundary theory we have defined is one where the time function is reparametrization invariant. This is seen by the fact that the momenta \eqref{eq:HJ momenta} obey quadratic identities \eqref{eq:quad id} and, thus, only pick out \emph{directions} and not absolute speeds on configuration space. The fact that $S$ is only invariant under dilatations if the time variable is appropriately rescaled, implies that, in the dual theory, invariance under dilatations must be accompanied by an appropriate reparametrization. This property can be seen, in the scale-invariant theory defined by the Hamiltonian constraint
\begin{equation}
	H = p^2 - \frac {m^2 \ell^2}{q^2},
\end{equation}
because the generator of dilatations
\begin{equation}
	D = p \cdot q
\end{equation}
only commutes weakly with the Hamiltonian constraint
\begin{equation}
	\pb D H = 2H\,.
\end{equation}
Thus, a dilatation is only preserved by the time evolution for a given parametrization.

One final point is that the logarithmically divergent term 
\begin{equation}
	 \sfrac {m\ell}2 \ln \lf( \frac {x\in^- x\out^-} {\ell^2} \rt)
\end{equation}
in \eqref{eq:boundary S} could be regulated by adding an appropriate counter term to the action, but this counter term will necessarily break conformal invariance. This is, again, because the transformation properties of $x^-$ are necessary for the invariance of $S$ under dilatations and special conformal transformations. The quantity $\frac {m\ell}2$ is, thus, somewhat reminiscent of a conformal anomaly. It would be interesting to investigate, in a quantum field theoretic generalization of this model, whether this divergence can indeed be related to a conformal anomaly in the boundary CFT.

\subsection{Physical Interpretation}

The procedure of the previous section outlines an exact duality between two classical systems that are mathematically equivalent. This duality is holographic in a sense very similar to the AdS/CFT correspondence \cite{Maldacena:ads_cft,Witten} --- although here it is more like a dS/CFT correspondence \cite{Strominger:holo_cosmo,Skenderis:holo_uni} --- in that, on one side of the duality, there is a spacetime theory and, on the other side, there is a boundary theory invariant under conformal transformations. The spacetime we have chosen is maximally symmetric, so that our model is much simpler than most standard holographic models, and the bulk ismoetries map to conformal transformations of the boundary theory.

Because we have a precise and exact duality, we can ask about the \emph{physical} relationship between these two theories. Unfortunately, any physical connection between the two theories appears to be extremely non-local. This comes from the physical interpretation of the coordinates $X^i$ on the future versus the past boundary. Between these boundaries, the stereographic projection of a point goes to the projection of its antipodal point. Because, as was already discussed, only null geodesics can traverse exactly half the sky, it is ensured that the proper time between events with asymptotic coordinates $X^i\in = X^i\out$ is zero. Thus, a particle at rest in the dual theory corresponds to a null geodesic in the bulk. Consequently, the \emph{faster} a particle moves in the bulk theory, the \emph{slower} it moves in the boundary theory. Clearly, the trajectories of particles in both theories vary dramatically. It is only the asymptotic configurations that should match up and, even then, only with their antipodal points. In general, the evolution of the particles in the dual theory is determined by the \emph{entire} evolution of the bulk system.

The interpretation of the symmetries of both theories can also be investigated. In the bulk theory, the spacetime isometries are conventionally attributed to some canonical choice of rods and clocks for physical reference frames moving differently with respect to each other. For example, otherwise identical reference frames moving with different velocities correspond to boosted observers. In the dual boundary theory, the bulk isometries are mapped to conformal symmetries of the Hamilton--Jacobi functional. In the case of dilatations and special conformal transformations, there is an accompanying time reparametrization. However, these time reparametrizations are global and, consequently, there is only a single notion of simultaneity. Instead, relativity of simultaneity has been replaced with a relative notion of scale. In the boundary theory, we have \emph{conformal} reference frames that agree on a global notion of time, but disagree about their conventional choice of scale. This interchangeability of the notions of relative time and relative scale is an important feature both of holographic dualities --- where it is manifest as \emph{holographic renormalization} --- and shape dynamics --- where it is appears through the \emph{symmetry trading} of local hypersurfaces deformation and local Weyl transformations. The physical mechanism behind this is very interesting and will constitute the subject of future investigations.

\section{Shape Dynamics and Generalizations}\label{sec:SD and gen}

In this section, we will briefly comment on the potential relationship between the duality presented here and the duality between General Relativity and Shape Dynamics. In both dualities, there exists a notion of interchangeability between relative time and relative scale. However, while the later is a spatially local duality, in that it involves symmetry transformations that are independent at different points in space, the former is purely global: the spaces considered are homogeneous and the symmetries transformations are global isometries. This suggests that there might exist a natural generalization of this model which promotes the global symmetries to local ones.

Cartan geometry \cite{Sharpe:Cartan_geometry} is a natural mathematical tool that could be used for performing this generalization. The idea behind it is to use a \emph{model geometry}, which can be any homogeneous space, to build general curved geometries that look locally like the homogeneous model space chosen. In the case of GR, dS spacetime itself can be used a model space (see \cite{Wise:MM_paper} for a description of how this possible). However, the asymptotic coordinates $X^i$ are representations of the projective light-cone (as described above), which represents a homogeneous space of its own. Perhaps there is a way to describe Shape Dynamics in terms of a Cartan geometry modelled off the projective light-cone? Some progress has already been made in this direction \cite{gryb:2_plus_1,Wise:2013uma}, but several obstacles still remain. If this program were successful, then the present work could illustrate how it might be possible to construct local reference frames for Shape Dynamics by relating them to local reference frames in GR, which are well understood. However, the holographic nature of the correspondence presented here seems incompatible with the duality between Shape Dynamics and GR. Although this duality is non-local in \emph{space}, it does not appear to be non-local in \emph{time} also, in contrast to the holographic duality presented here. This would seem to suggest that there may be a key insight missing in connecting Shape Dynamics to GR through Cartan geometry.

The main difficulty presents itself in finding an exact representational equivalence between the conformal and spacetime symmetries. Insisting on this in the way we have approached the problem here has led directly to a holographic picture instead of a bulk-bulk duality. This is because we were able to make use of the shape freezing near the conformal boundaries of dS to establish our correspondence. In the bulk, boosts invariably change the definition of simultaneity making the notion of instantaneous shape ambiguous. We are then led to conclude that, without some further insight, it may be more appropriate to think of the relationship between conformal symmetries and hypersurface deformations more as ``complimentary'' or ``hidden'' symmetries of gravity, rather than strictly ``dual''.

Other generalizations of our model are possible. A straightforward thing to do would be to add interactions of the bulk theory to see how this might effect the Hamiltonian of the dual theory. A slightly more ambitious, but still very manageable, generalization would be to consider matter fields in the bulk. A bulk scalar field would be extremely interesting to study in the dual framework. Quantization is a further option. In this case, the role of $\hbar$, which we have completely ignored in our discussions so far, could be studied directly and compared with the usual results known from gauge/gravity dualities. Finally, lifting the condition of homogeneity of the spacetime could be explored (this is notwithstanding the Cartan geometric methods discussed above). The main difficulty for our context here is that asymptotic dS spacetimes exhibit a supertranslation ambiguity when appropriate fall-off conditions are imposed on the spacetime metric. This allows for a larger, infinite dimensional symmetry group that presents additional complications.

\section{Conclusions}

We have presented a model, inspired by the duality between Shape Dynamics and General Relativity, where a bulk theory of free particles in dS spacetime can be mapped to a reparametrization and conformally invariant theory on the conformal boundaries of dS. The bulk dS ismoetries map to conformal symmetries in the dual theory. This map is interpreted as a correspondence between bulk inertial reference frames and boundary conformal reference frames, who only see the scale-invariant information about the instantaneous \emph{shape} of the particle system. This leads to a definition of \emph{Shape Observers} who are dual to inertial observers in the spacetime picture. The correspondence is holographic and reminiscent of the AdS/CFT correspondence, where the bulk asymptotic form of the on-shell action is used to define the Hamilton--Jacobi functional of the dual conformal theory. Many different generalizations have been suggested that could incrementally lead to a deeper understanding of holographic dualities. The point to emphasise is that the correspondence is both precise and exact, so that both sides of the duality can be worked out explicitly. Although the duality presented shares many important features with the Shape Dynamics/GR duality --- such as the interchangeability of relative scale and relative time --- the holographic nature of the duality, which does not seem to be a necessary feature of Shape Dynamics, suggests that there are still key insights that are missing in order to link the two approaches. There seems to be more work to be done before connecting this bottom-up approach to a full explanation of the connection, alluded to in the introduction, between the physical role played by spacetime general covariance in General Relativity and the role of spatial general covariance and local scale invariance in Shape Dynamics. This will be the subject of future investigations.

\section*{Acknowledgements}

Many thanks to Josh Cooperman and, particularly, Derek Wise for many helpful discussions. Thanks also to Marc Ngui for providing concepts for the illustrations and to Fabries van den Heuvel for many useful exchanges and for double checking some calculations. This work was supported by the Natural Science and Engineering Research Council (NSERC) of Canada and by the Netherlands Organisation for Scientific Research (NWO) (Project No. 620.01.784).

\clearpage

\appendix

\section{Physical Clock and Rod Readings from Reference Frames} \label{apx:ref frames}

A reference frame is formally represented by a congruence whose worldlines are the integral curves of a time-like vector field $u^\mu$ in a spacetime manifold $\mathcal M$. In this section, we briefly sketch an explicit construction for extracting a particular set of physical clock and rod readings --- represented by the scalar fields $X^a$, where $a = 1,\hdots, 4$ --- and the conditions under which this can be done. This is meant to be a proof of principle to illustrate one way in which such a decomposition could be done. Such a construction is certainly not unique and, moreover, we do not even claim that there is a simple physical interpretation for our decomposition. For a different decomposition in terms of ``velocity potentials'' motivated by analogies with fluid dynamics, see \cite{schutz:1970}.

For our construction, we are interested in physical reference frames that a realistic local observer could use to make time and positions readings for events in some spacetime region. We will thus make the physical assumption that the boundary conditions of the spacetime region of interested are such that one can apply a Hodge decomposition to the vector field $u^\mu$. Technically, this requires that the region be defined on a Sobolev space, but practically this restriction is physically mild. In Minkowski space, the Hodge decomposition is nothing more than the usual Hemholtz decomposition of a vector field. A second requirement for the Hodge decomposition is the existence of a spacetime metric. We therefore need to assume some alternative means of measuring the spacetime metric in order to apply our construction.

Given these requirements, we can use the spacetime metric to construct the $1$-form $u = u_\mu \otimes dx^\mu$, where $u_\mu = g_{\mu\nu} u^\nu$. Then, the (unique) Hodge decomposition takes the form
\begin{equation}
	u = u^\parallel + u^\perp\,,
\end{equation}
where $u^\parallel = \delta \alpha $, $u^\perp = \de \beta$, $\de$ is the exterior derivative, and $\delta$ is the co-differential defined by $\delta = (-1)^k \star^{-1}\, \de\, \star$, where $k$ is the grading of the differential form it acts on and $\star$ is the Hodge product.\footnote{Given certain global conditions, $u^\parallel$ may also contain a harmonic function $\gamma$ such that $\Delta \gamma = 0$, where $\Delta$ is the Laplacian.} The co-differential involves the use of the Hodge product and thus requires knowledge of the spacetime metric. Because $u^\perp$ is an exact form, it is hypersurface orthogonal by Frobenius' theorem. The hypersurfaces, $\Sigma$, it defines label constant values of the scalar $\beta$, which we can identify as the time $X^4 = \beta$. The pullback of $u^\parallel$ onto $\Sigma$ can be used to define a new 1-form $v$ on $\Sigma$. Now, a Hodge decomposition can be similarly performed on $v$ giving
\begin{equation}
	v = v^\parallel + v^\perp\,,
\end{equation}
where $v^\parallel = \delta \sigma $ and $v^\perp = \de \rho$. Again, because $v^\perp$ is hypersurface orthogonal, it can be used to foliate $\Sigma$ by constant $\rho$ surfaces. This can be used to define one of the spatial coordinate scalars $X^1 = \rho$. We can proceed in a similar fashion, taking pullbacks onto hypersurfaces and performing the Hodge decomposition, to construct the remaining two scalars $X^2$ and $X^3$. This completes our construction.

The requirements for performing this decomposition are now clear. We need: i) the appropriate boundary conditions for applying the Hodge decomposition, ii) knowledge of the spacetime metric, iii) that at no stage of the process the perpendicular component of the decomposed vector field be zero. The last requirement simply means that the vector field $u^\mu$ must have its maximum number of independent components so that 4 independent scalars can be formed from it. A nice example that concretely implements a decomposition of this form is the Gaussian reference fluid of Kucha\v r and Torre presented in \cite{Kuchar:1990vy}.

\section{Timelike Geodesics in de~Sitter}\label{apx:geodesics}

In this appendix, we compute the timelike geodesics for dS spacetime by extremizing the proper length, $S$, along some trajectory $\dot x^\mu(t)$
\begin{equation}\label{action 1 part}
	S = \int_{x^\mu(t_1)}^{x^\mu(t_2)}  \de t \lf[ {\textstyle \frac 1 2} m\sqrt{ - \eta_{\mu\nu} \dot x^\mu \dot x^\nu } + \lambda \lf( \eta_{\mu\nu} x^\mu x^\nu - \ell^2 \rt) \rt]\,
\end{equation}
between two points $x^\mu(t_1)$ and $x^\mu(t_2)$, where all coordinates, $x^\mu$, are defined in the ambient $\mathbbm R^{\De+1,1}$. The Lagrange multiplier $\lambda$ enforces the constraint that the trajectory remain on the dS hyperboloid. This can be taken as the action, $S$, for some point particle of mass $m$ in dS.

Hamiltonian methods are particularly convenient for dealing with constrained systems of the form \eqref{action 1 part}, by making use of the Dirac algorithm (see \cite{ henneaux_teit:quant_gauge} for a description). We will begin the construction of the Hamiltonian by defining the momenta
\begin{align}
	p_\mu &= \diby S{\dot q^\mu} = \frac{- m \,\eta_{\mu\nu} \dot x^\nu}{\sqrt{ - \eta_{\mu\nu} \dot x^\mu \dot x^\nu  }}\,, & p_\lambda &= \diby S{\dot \lambda} = 0 \,,
\end{align}
which obey the primary constraints
\begin{align}
	\ham &\equiv \eta^{\mu\nu} p_\mu p_\nu + m^2 \approx 0\,, & p_\lambda & \approx 0 \,.
\end{align}
The second constraint is second class wrt the Hamiltonian $H = p_\mu \dot q^\mu + p_\lambda \dot \lambda - S$ and leads to the secondary constraint
\begin{equation}
	G = \eta_{\mu \nu} x^\mu x^\nu - \ell^2 \approx 0\,.
\end{equation}
After a simple (and standard) redefinition of the Lagrange multiplier $\lambda$ and using the lapse multiplier, $N$, the resulting Hamiltonian is
\begin{equation}
	H = N\lf( \eta^{\mu\nu} p_\mu p_\nu + m^2 \rt) + \lambda \lf( \eta_{\mu\nu} x^\mu x^\nu - \ell^2 \rt).
\end{equation}
However, this is still \emph{not} the total Hamiltonian of the system because these constraints are not first class:
\begin{equation}
	\pb{ x^2 - \ell^2}{p^2 + m^2} = 4 x\cdot p\,,
\end{equation}
where we have used the abbreviations $x^2 \equiv \eta_{\mu\nu} x^\mu x^\nu$, $p^2 \equiv \eta^{\mu\nu} p_\mu p_\nu$, and $x\cdot p = x^\mu p_\mu$. We, thus, obtain the further constraint
\begin{equation}
	D \equiv x^\mu p_\mu \approx 0\,,
\end{equation}
which, in turn, is second class wrt both primary constraints:
\begin{align}
	\pb \ham D &= \pb {p^2 + m^2}{q\cdot p} = -2p^2 \approx -2m^2 \\
	\pb G D &= \pb {x^2 - \ell^2}{q\cdot p}  = 2x^2 \approx 2\ell^2.\label{pb G D}
\end{align}

An efficient way to treat this second class system is to define the Dirac bracket for one of the second class pairs and then use the remaining constraint as the first class Hamiltonian, computing Hamilton's equations using the appropriate Dirac bracket for the second class system. To do this, we will first construct the Dirac bracket for the second class pair $(G,D)$. The Poisson bracket \eqref{pb G D} has the trivial inverse $\frac 1 {2\ell^2}$. We can then use the standard definition of the Dirac bracket to obtain
\begin{equation}
	\pb f g_\text{Db} = \pb f g - \pb f {x^2 - \ell^2}{\textstyle \frac 1 {2\ell^2} } \pb { x\cdot p } g - (f \leftrightarrow g),
\end{equation}
for two arbitrary phase space functions $f$ and $g$. This leads to the modified symplectic structure
\begin{align}
	\pb {x^\mu}{p_\nu}_\text{Db} &= \delta^\mu_\nu + \frac 1 {2\ell^2} \pb{x^\mu}{ x\cdot p} \pb{x^2 - \ell^2}{p_\nu} \\
	                             &= \delta^\mu_\nu + \frac{x^\mu x_\nu}{\ell^2}\,
\end{align}
and
\begin{align}
	\pb {p_\mu}{p_\nu}_\text{Db} &= -\frac 1 {2\ell^2} \lf[ \pb{p_\mu}{x^2 - \ell^2} \pb{x\cdot p}{p_\nu} - \pb{p_\mu}{x\cdot p} \pb{x^2 - \ell^2}{p_\nu}\rt]\\
	                             &= \frac 1 {\ell^2} \lf[  x_\mu p_\nu - p_\mu x_\nu  \rt]\\
	                             &= 2\frac{x_{[\mu} p_{\nu]}}{\ell^2}\,.
\end{align}
Also, $\pb{x^\mu}{x^\nu} = 0$.

Using this new symplectic structure, we can use the strong equations $x^2 = \ell^2$ and $x\cdot p = 0$ then work out Hamilton's equations for the first class Hamiltonian
\begin{equation}
	H = N (p^2 + m^2)\,.
\end{equation}
These are:
\begin{align}
	\dot x^\mu &= \pb{x^\mu}{H}_\text{Db} \\
			   &= 2N \lf( p^\mu + x^\mu \frac{x\cdot p}{\ell^2}\rt)\\
			   &= 2N p^\mu\,,
\end{align}
and
\begin{align}
	\dot p_\mu &= \pb{p_\mu}{H}_\text{Db} \\
			   &= \frac {2N}{\ell^2} \lf[ x_\mu p^2 - p_\mu x\cdot p \rt] \\
			   &= \frac {2Nm^2}{\ell^2} x_\mu\,.
\end{align}
They combine to give
\begin{equation}\label{eq:geo eoms}
	\frac{\de^2 x^\mu}{\de\tau^2} = \frac {m^2}{\ell^2} x^\mu,
\end{equation}
where $\de\tau = 2N \de t$ is the proper time along the trajectory.

Solutions of these equations of motion must solve the initial data constraint $p^2 + m^2 = 0$ on top of the strong equations $x^2 = \ell^2$ and $x\cdot p = 0$. The last part of the gauge invariance is the reparametrization invariance of the curve, which can be fixed by restricting to a proper time parametrization. In term of the proper time, the most general solution to \eqref{eq:geo eoms} is
\begin{equation}\label{bulk soln}
	\boxed{x^\mu(\tau) = A^\mu e^{\sfrac m \ell \tau } + B^\mu e^{-\sfrac m \ell \tau }}\,,
\end{equation}
for some integration constants $A^\mu$ and $B^\mu$ that we will interpret shortly.
\begin{equation}\label{bulk dot x}
	p_\mu = \dot x^\mu(\tau) = \frac m \ell \lf[ A^\mu e^{\sfrac m \ell \tau } - B^\mu e^{-\sfrac m \ell \tau } \rt]\,.
\end{equation}
Note that other parametrizations are, of course, possible. This will only change the relations between and interpretation of the integration constants.

The general solutions here must obey the initial value constraints and the gauge fixing conditions. These simply put restrictions on the integration constants, which can easily be worked out. If we first consider the initial value constraint $p^2 + m^2 = 0$ and the strong equation $x^2 - \ell^2 = 0$, we find that the combination
\begin{equation}
	\frac {\ell^2}{m^2} \lf( p^2 + m^2 \rt) - \lf( x^2 - \ell^2 \rt) = 0
\end{equation}
implies that
\begin{equation}
	\boxed{2A\cdot B = \ell^2}\,.
\end{equation}
Similarly, the combination
\begin{equation}
	\frac {\ell^2}{m^2} \lf( p^2 + m^2 \rt) + \lf( x^2 - \ell^2 \rt) = 0
\end{equation}
implies
\begin{equation}
	A^2 e^{2\sfrac m \ell \tau} + B^2 e^{-2\sfrac m \ell \tau} = 0\,.
\end{equation}
Using the fact that the additional strong equation
\begin{equation}
	x\cdot p = 0
\end{equation}
implies
\begin{equation}
	A^2 e^{2\sfrac m \ell \tau} - B^2 e^{-2\sfrac m \ell \tau} = 0\,,
\end{equation}
we immediately find
\begin{equation}
	\boxed{A^2 = B^2 = 0}\,.
\end{equation}
Thus, $A^\mu$ and $B^\mu$ are two null vectors normalized such that $2A\cdot B = \ell^2$. Because of the Lorentzian signature, this normalization requires that $A^\mu$ and $B^\mu$ be pointing in opposite directions in time. We can, thus, pick $A^\mu$ to be `backward pointing' while $B^\mu$ is `forward pointing'. Furthermore, since as $t\to +\infty$, $x^\mu \to A^\mu$ and as $t \to -\infty$, $x^\mu \to B^\mu$, these null vectors represent `ingoing', $\xi\in^\mu$ and `outgoing', $\xi^\mu\out$, directions on the $\De$-sphere at past and future null infinity, respectively. See Figure~\ref{fig:geodesics} in the text for a visual representation of these solutions.

\clearpage
\bibliographystyle{utphys}
\bibliography{mach}

\providecommand{\href}[2]{#2}\begingroup\raggedright\begin{thebibliography}{10}

\bibitem{einstein:gen_rel}
A.~Einstein, ``{The foundation of the general theory of relativity},''
  \href{http://dx.doi.org/10.1002/andp.200590044}{{\em Annalen Phys.}
  {\bfseries 49} (1916) 769--822}.

\bibitem{kretschmann:gen_cov}
E.~Kretschmann, ``{$\ddot{\text{U}}$ber den physikalischen Sinn der
  Relativit$\ddot{\text{a}}$tspostulate, A. Einsteins neue und seine
  urspr$\ddot{\text{u}}$ngliche Relativit$\ddot{\text{a}}$tstheorie},'' {\em
  Annalen der Physik} {\bfseries 53} (1917) 575.

\bibitem{Westman:2007yx}
H.~Westman and S.~Sonego, ``{Coordinates, observables and symmetry in
  relativity},'' \href{http://dx.doi.org/10.1016/j.aop.2009.03.014}{{\em Annals
  Phys.} {\bfseries 324} (2009) 1585--1611},
\href{http://arxiv.org/abs/0711.2651}{{\ttfamily arXiv:0711.2651 [gr-qc]}}.
%%CITATION = ARXIV:0711.2651;%%.

\bibitem{Kuchar:1990vy}
K.~V. Kuchar and C.~G. Torre, ``{Gaussian reference fluid and interpretation of
  quantum geometrodynamics},''
\href{http://dx.doi.org/10.1103/PhysRevD.43.419}{{\em Phys.Rev.} {\bfseries
  D43} (1991) 419--441}.
%%CITATION = PHRVA,D43,419;%%.

\bibitem{Dieks:frame_equiv}
D.~Dieks, ``Another look at general covariance and the equivalence of reference
  frames,'' {\em Studies in History and Philosophy of Science Part B}
  {\bfseries 37} no.~1, (2006) 174--191.

\bibitem{gryb:shape_dyn}
H.~Gomes, S.~Gryb, and T.~Koslowski, ``{Einstein gravity as a 3D conformally
  invariant theory},''
  \href{http://dx.doi.org/10.1088/0264-9381/28/4/045005}{{\em Class. Quant.
  Grav.} {\bfseries 28} (2011) 045005},
  \href{http://arxiv.org/abs/1010.2481}{{\ttfamily arXiv:1010.2481 [gr-qc]}}.

\bibitem{Gomes:linking_paper}
H.~Gomes and T.~Koslowski, ``{The Link between General Relativity and Shape
  Dynamics},'' {\em Class.Quant.Grav.} {\bfseries 29} (2012) 075009,
  \href{http://arxiv.org/abs/1101.5974}{{\ttfamily arXiv:1101.5974 [gr-qc]}}.

\bibitem{JuliansReview}
J.~Barbour, ``{Shape Dynamics. An Introduction},''
  \href{http://arxiv.org/abs/1105.0183}{{\ttfamily arXiv:1105.0183}}.

\bibitem{gryb:thesis}
S.~Gryb, {\em Shape dynamics and Mach's principles: Gravity from conformal
  geometrodynamics}.
\newblock PhD thesis, University of Waterloo, 2011.
\newblock \href{http://arxiv.org/abs/1204.0683}{{\ttfamily arXiv:1204.0683
  [gr-qc]}}.

\bibitem{Hooft:1993gx}
G.~'t~Hooft, ``{Dimensional reduction in quantum gravity},''
\href{http://arxiv.org/abs/gr-qc/9310026}{{\ttfamily arXiv:gr-qc/9310026
  [gr-qc]}}.
%%CITATION = GR-QC/9310026;%%.

\bibitem{Susskind:1994vu}
L.~Susskind, ``{The World as a hologram},''
  \href{http://dx.doi.org/10.1063/1.531249}{{\em J.Math.Phys.} {\bfseries 36}
  (1995) 6377--6396},
\href{http://arxiv.org/abs/hep-th/9409089}{{\ttfamily arXiv:hep-th/9409089
  [hep-th]}}.
%%CITATION = HEP-TH/9409089;%%.

\bibitem{Wise:2013uma}
D.~K. Wise, ``{Holographic Special Relativity},''
\href{http://arxiv.org/abs/1305.3258}{{\ttfamily arXiv:1305.3258 [hep-th]}}.
%%CITATION = ARXIV:1305.3258;%%.

\bibitem{Maldacena:ads_cft}
J.~M. Maldacena, ``{The large N limit of superconformal field theories and
  supergravity},'' \href{http://dx.doi.org/10.1023/A:1026654312961}{{\em Adv.
  Theor. Math. Phys.} {\bfseries 2} (1998) 231--252},
  \href{http://arxiv.org/abs/hep-th/9711200}{{\ttfamily arXiv:hep-th/9711200}}.

\bibitem{Witten}
E.~Witten, ``{Anti-de Sitter space and holography},'' {\em Adv. Theor. Math.
  Phys.} {\bfseries 2} (1998) 253--291,
  \href{http://arxiv.org/abs/hep-th/9802150}{{\ttfamily arXiv:hep-th/9802150}}.

\bibitem{Strominger:holo_cosmo}
A.~Strominger, ``{Inflation and the dS / CFT correspondence},'' {\em JHEP}
  {\bfseries 0111} (2001) 049,
  \href{http://arxiv.org/abs/hep-th/0110087}{{\ttfamily arXiv:hep-th/0110087
  [hep-th]}}.

\bibitem{Skenderis:holo_uni}
P.~McFadden and K.~Skenderis, ``{The Holographic Universe},''
  \href{http://dx.doi.org/10.1088/1742-6596/222/1/012007}{{\em
  J.Phys.Conf.Ser.} {\bfseries 222} (2010) 012007},
  \href{http://arxiv.org/abs/1001.2007}{{\ttfamily arXiv:1001.2007 [hep-th]}}.

\bibitem{Hehl:1994ue}
F.~W. Hehl, J.~D. McCrea, E.~W. Mielke, and Y.~Ne'eman, ``{Metric affine gauge
  theory of gravity: Field equations, Noether identities, world spinors, and
  breaking of dilation invariance},''
  \href{http://dx.doi.org/10.1016/0370-1573(94)00111-F}{{\em Phys.Rept.}
  {\bfseries 258} (1995) 1--171},
\href{http://arxiv.org/abs/gr-qc/9402012}{{\ttfamily arXiv:gr-qc/9402012
  [gr-qc]}}.
%%CITATION = GR-QC/9402012;%%.

\bibitem{MacDowell:1977jt}
S.~MacDowell and F.~Mansouri, ``{Unified Geometric Theory of Gravity and
  Supergravity},''
\href{http://dx.doi.org/10.1103/PhysRevLett.38.1376,
  10.1103/PhysRevLett.38.739}{{\em Phys.Rev.Lett.} {\bfseries 38} (1977) 739}.
%%CITATION = PRLTA,38,739;%%.

\bibitem{Stelle:1979va}
K.~Stelle and P.~C. West, ``{DE SITTER GAUGE INVARIANCE AND THE GEOMETRY OF THE
  EINSTEIN-CARTAN THEORY},''
\href{http://dx.doi.org/10.1088/0305-4470/12/8/003}{{\em J.Phys.} {\bfseries
  A12} (1979) L205--L210}.
%%CITATION = JPHGB,A12,L205;%%.

\bibitem{Wise:MM_paper}
D.~K. Wise, ``{MacDowell-Mansouri gravity and Cartan geometry},''
  \href{http://dx.doi.org/10.1088/0264-9381/27/15/155010}{{\em
  Class.Quant.Grav.} {\bfseries 27} (2010) 155010},
  \href{http://arxiv.org/abs/gr-qc/0611154}{{\ttfamily arXiv:gr-qc/0611154
  [gr-qc]}}.

\bibitem{Randono:2010cq}
A.~Randono, ``{Gauge Gravity: a forward-looking introduction},''
\href{http://arxiv.org/abs/1010.5822}{{\ttfamily arXiv:1010.5822 [gr-qc]}}.
%%CITATION = ARXIV:1010.5822;%%.

\bibitem{Sharpe:Cartan_geometry}
R.~W. Sharpe, {\em Differential Geometry: Cartan's Generalization of Klein's
  Erlangen Program}.
\newblock Springer, New York, 1997.

\bibitem{gryb:2_plus_1}
S.~Gryb and F.~Mercati, ``{2+1 gravity on the conformal sphere},''
  \href{http://dx.doi.org/10.1103/PhysRevD.87.064006}{{\em {Phys. Rev. D}}
  {\bfseries {87}} (2013) {064006}},
  \href{http://arxiv.org/abs/1209.4858}{{\ttfamily arXiv:1209.4858 [gr-qc]}}.

\bibitem{schutz:1970}
B.~Schutz, ``Perfect fluids in general relativity: Velocity potentials and a
  variational principle,'' {\em Phys. Rev.} {\bfseries D2} no.~12, (1970) 2762
  -- 2773.

\bibitem{henneaux_teit:quant_gauge}
M.~Henneaux and C.~Teitelboim, {\em {Quantization of gauge systems}}.
\newblock Univ. Pr., Princeton, USA, 1992.
\newblock 520 p.

\end{thebibliography}\endgroup

\end{document}